\documentclass[11pt]{article}

\usepackage[margin=1in]{geometry}
\usepackage{graphicx}    
\usepackage{amsmath,amssymb}
\usepackage{setspace}
\usepackage{booktabs}
\usepackage{siunitx}    
\usepackage{url}
\usepackage[colorlinks=true,linkcolor=black,citecolor=black,urlcolor=black]{hyperref}
\usepackage{placeins}       
\usepackage{paralist}       
\usepackage{threeparttable} 

\newcommand{\citep}[2][]{\cite{#2}}

\title{Generative AI in Sociological Research:\\ State of the Discipline}

\author{%
AJ Alvero\thanks{Center for Data Science for Enterprise and Society, Cornell University. Corresponding author: \texttt{aja326@cornell.edu}. All authors contributed equally. We thank all the respondents of our survey for being generous with their time. We are indebted to Kim Weeden and Cat Dang Ton for giving us crucial comments on an early version of this paper. We are also grateful for the important feedback we received at the \textit{ASA Session on Culture and Computational Social Science}, the \textit{Sociological Science Conference} at Cornell University, the \textit{International Network of Analytical Sociology Conference} at Columbia University, the \textit{Institute for Analytical Sociology Symposium} at Link\"oping University, and the \textit{Culture and Action Network} at the University of Chicago.}%
\and Dustin S.~Stoltz\thanks{Department of Sociology and Anthropology, Lehigh University}%
\and Oscar Stuhler\thanks{Department of Sociology, Northwestern University}%
\and Marshall A.~Taylor\thanks{Department of Sociology, New Mexico State University}%
}

\date{}

\begin{document}
\maketitle

\begin{abstract}
\singlespacing \noindent 
Generative artificial intelligence (GenAI) has garnered considerable attention for its potential utility in research and scholarship. A growing body of work in sociology and related fields demonstrates both the potential advantages and risks of GenAI, but these studies are largely proof-of-concept or specific audits of models and products. We know comparatively little about how sociologists actually use GenAI in their research practices and how they view its present and future role in the discipline. In this paper, we describe the current landscape of GenAI use in sociological research based on a survey of authors in 50 sociology journals. Our sample includes both computational sociologists and non-computational sociologists and their collaborators. We find that sociologists primarily use GenAI to assist with writing tasks: revising, summarizing, editing, and translating their own work. Respondents report that GenAI saves time and that they are curious about its capabilities, but they do not currently feel strong institutional or field-level pressure to adopt it. Overall, respondents are wary of GenAI's social and environmental impacts and express low levels of trust in its outputs, but many believe that GenAI tools will improve over the next several years. We do not find large differences between computational and non-computational scholars in terms of GenAI use, attitudes, and concern; nor do we find strong patterns by familiarity or frequency of use. We discuss what these findings suggest about the future of GenAI in sociology and highlight challenges for developing shared norms around its use in research practice.
\end{abstract}

\doublespacing

\section{Introduction}

Following the release of the Generative Pre-trained Transformer-3 (GPT-3) in 2020 and the browser-based conversational interface ChatGPT in 2022, academics have been forced to grapple with generative artificial intelligence. For better or worse, products and tools that fall under the label ``GenAI'' are starting to make inroads into the major aspects of academic life: classroom instruction \citep{Xie2025-ze}, peer review \citep{Li2024-bu}, administration \citep{An2025-rr}, and not least \textit{research} \citep{Broska2024-od, Bail2024-pa, zhang2025exploring}. 

Thus far, the use of GenAI as a tool for sociological research has mainly been explored from the perspective of research products. Research articles offer ``proof of concepts'' by identifying important use cases and limitations (see, e.g., \citep{davidskar24, Boelaert2025-la}). Meanwhile, less attention has been given to how and why scholars in our discipline decide to use or avoid GenAI \citep{Watermeyer2024-kp, Watermeyer2024-pt}. Additionally, while there have been a series of recent publications that showcase high-level applications of GenAI---say, classifying images \citep{Law2024-jn} or extracting information from text \citep{Stuhler2025-rp}---there has been comparatively little open discussion about how sociologists may use GenAI for more every-day aspects of research -- be it for writing, generating ideas, debugging code, or otherwise.

In this paper, we present a survey for which we asked sociologists and their collaborators about how they perceive and use (or do not use) GenAI in their research. Our sampling frame includes the authors of all articles published in 50 sociology journals in the last five years. We contacted a random sample of these authors, as well as all authors who published ``computational'' articles. This allows us to paint the landscape of GenAI usage and attitudes among the population of researchers contributing to recent sociological scholarship. Going beyond speculation about what technology ``could'' do for sociologists, we offer the first systematic evidence on how GenAI is currently being used and seen by sociologists. Thereby, we provide an empirical baseline for a more grounded discussion on GenAI's role in sociological research.

To preview our results, the most common usage of GenAI is for writing assistance, especially for grammar, spelling, and paraphrasing sections of one's own writing. We find that roughly 34\% of sociologists and their collaborators have used GenAI in this capacity---on par with findings from surveys of scholars in other fields \citep{Kwon2025-ay, Ng2025-kf}. Respondents use GenAI because they perceive that it saves them time; out of curiosity; and because it is increasingly incorporated into tools they already use (e.g., search engines). However, very few reported feeling pressure to use GenAI from their collaborators, field, or institution. Although we anticipated differences in use and attitudes between those who use computational methods and those who do not, we found very few. Similarly, we find that expertise---in terms of self-reported familiarity and use frequency---is a weak predictor of attitudes. Generally, the vast majority of scholars are very concerned about the social and environmental consequences of GenAI and also distrust GenAI outputs. Finally, scholars agreed that GenAI would likely improve in the next few years, but were divided about whether it would have a net positive effect on the field.

\section{Background}

\subsection{What Is ``Generative'' AI?}

Language surrounding ``artificial intelligence'' is often imprecise, with a wide range of technologies grouped under this umbrella \citep{Bender2025-wa}. Therefore, we begin by briefly explaining how we define GenAI in this study (also see our primer on GenAI in email invitation sent to survey respondents in Appendix~\ref{app0}). 

The distinguishing characteristic of GenAI from other computational methods is in the name: as opposed to ``discriminative'' models that find optimal boundaries in data, primarily for classification tasks, generative models are designed to \textit{generate} text, images, audio, and video. We focus on text generation models, as that is also the focus of the current literature in sociology and represents the most common form of GenAI interaction \citep{Zhang2025-st}. When employing such models, users typically write a \textit{prompt} containing the text that a trained model will take as its point of departure, such as instructions, questions, or an arbitrary query. This prompt is encoded as tokens\footnote{Tokens are the basic inputs and outputs in natural language processing. In LLMs, tokens are words, combinations of words (including complete sentences), parts of words (e.g., word stems), and/or common non-word characters like spaces or dashes. See \citep{mielke2021between} for additional detail.}, then the underlying model generates a \textit{probable} continuation of the prompt by identifying the most likely \textit{next token}, defined in probabilistic terms. The model then incorporates that new token into another round of generation and continues until a stopping rule, such as a maximum number of tokens to generate, is satisfied. In other words, the prompt provides a starting point from which the model starts generating; for a more thorough introduction and discussion of practical aspects, see \citep{Chae2023-yb}.

\subsection{Generative AI in Academic Research}

Setting aside quality, many of the tasks that comprise academic research are highly amenable to GenAI precisely because they are text-based. However, systematic surveys on the uptake of and attitudes toward GenAI for research have been scarce. Some cross-disciplinary surveys suggest that scholars in the technical and life sciences have embraced GenAI more quickly and are more optimistic about its prospects than those in the humanities or social sciences \citep{andersen2025generative, oup_ai_survey_2024, HrycyshynEassom2025ExplanAItions}. Most studies report that younger or early career scholars use GenAI more frequently \citep{perkowski2024generative, dorta2024generative, van2023ai, andersen2025generative, HrycyshynEassom2025ExplanAItions, Kwon2025-ay}, and consider it more acceptable to do so \citep{Kwon2025-ay}---though one survey found the opposite \citep{oup_ai_survey_2024}. Some report that men use GenAI more frequently than women \citep{dorta2024generative, perkowski2024generative, chakravorti2025socialscientistsroleai}, while others find no gender differences \citep{andersen2025generative}. A recurring theme noted by several studies is that opinions about GenAI are very heterogeneous, and that there is far from a consensus on what kinds of use of GenAI are legitimate \citep{andersen2025generative, Kwon2025-ay}.

Differences in sampling strategies, fielding time, survey design, and even the assumed definition of ``AI'' make it difficult to compare results and lead to contrasting findings. For instance, a 2024 survey of researchers by the publisher Wiley reports that 81\% of respondents used ChatGPT \citep{HrycyshynEassom2025ExplanAItions}. A survey of PhD-holding economists working at European central banks fielded in the same year reports that fewer than half are using ``OpenAI’s ChatGPT, Google Gemini, Github Copilot, Meta’s LLaMa, Anthropic’s Claude, or another generative AI tool'' \citep{perkowski2024generative}. While these studies exhibit heterogeneity in questions and responses, one key takeaway is that researchers across a range of fields have started to incorporate GenAI into their workflow.

\subsection{Generative AI in Sociological Research}

Sociologists have also begun to point out potential use cases of GenAI models for research tasks \citep{Bail2024-pa, Davidson2024-mo}. Empirical studies have examined whether GenAI can increase research efficiency, for example, by generating survey questions \citep{Gotz2024-lf}; imputing missing data \citep{Kim2023-fn}; engaging in an exploratory dialogue with qualitative data \citep{Ibrahim2024-ml, Hayes2025-bn}; and classifying, annotating, and extracting information from text or images \citep{Gilardi2023-re, Law2024-jn, Stuhler2025-rp, Nelson2025-zb, Maranca2025-li, Schwitter2025-li, Lin2025-vk}. Another branch of scholarship has explored GenAI's potential for simulating human behavior \citep{Kozlowski2025-qe, Anthis2025-xh, Broska2024-od, Alvero2024-iq}. Much of this work has focused on simulating responses to survey items \citep{Boelaert2025-la, Kim2023-fn, Broska2025SMR, Kozlowski2024-co, Zhang2025-st}; outside of sociology, see, e.g., \citep{Argyle_2023}, but GenAI can also be used to simulate human interaction and communication \citep{Karell2024-qn, Horton2023-do, argyle2023leveraging}. 

Sociologists have also pointed out how GenAI complicates sociological analysis. This includes homogenization of outputs \citep{Zhang2025-st, Alvero2024-iq}; injecting new forms of machine bias \citep{Boelaert2025-la, Stuhler2025-rp, Rister-Portinari-Maranca2025-dz}; and LLMs parroting government-approved ideological positions \citep{Waight2024-tj} or potentially the positions favored by the organizations producing the models \citep{Martin2023-mk}. In some cases, GenAI may perform only on par with, or even worse than, traditional methods that are often more transparent and less computationally costly \citep{Mu2023-rg, Nelson2025-zb, Stuhler2025-rp, Ashwin2025-af}. Furthermore, if sources of empirical data, such as open-ended survey responses \citep{Zhang2025-st} or social media posts \citep{sourati2025shrinking} are increasingly produced using GenAI, future research will be confronted with difficult questions about sociological explanation and inference.

In the midst of both excitement and skepticism, we are at a critical moment where GenAI is clearly making inroads into research practices, and not just for those who do ``computational'' work. Yet we have little systematic knowledge about how sociologists and their collaborators currently use GenAI, and about the field's sentiment towards this new technology. Our goal is to help close this gap.

\section{Data and Methods}

\subsection{Sample}

We used a bibliometric multistage sampling design to construct a representative sample of both computational and non-computational sociologists and their collaborators. The sample includes 219 non-computational authors and an over-sample of 214 computational authors (defined as those authoring an article with computational terms in the title, abstract, and keywords) for a total of 433 respondents. We used rake weights \citep{debell2009computing} to adjust for this oversampling as well as selection bias into our sample---namely, gender and location. Our sampling method is detailed in Appendix~\ref{app3}. Figure \ref{fig:schem} is a summary schematic of the sampling strategy.

\begin{figure}[!h]
\centering\includegraphics[width=.8\linewidth]{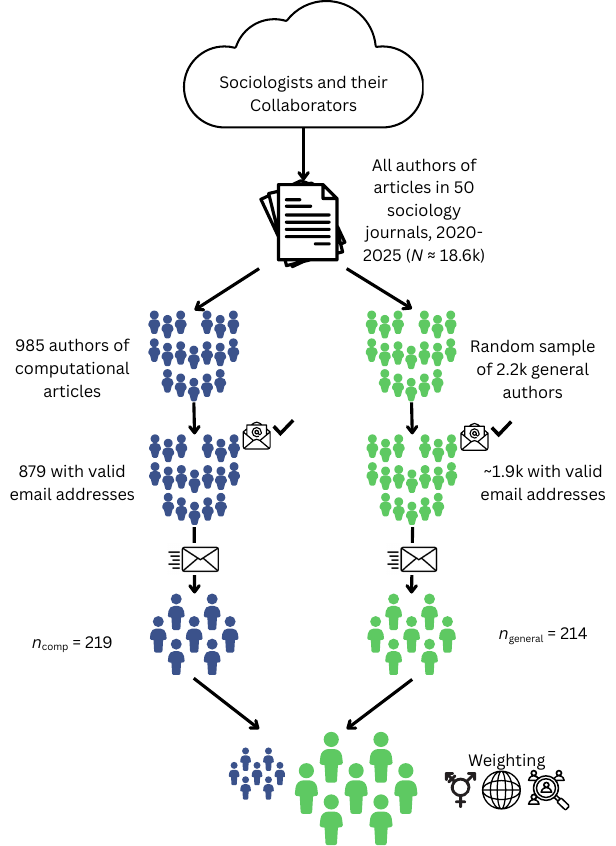}
\caption{Schematic Representation of Sampling Strategy}
\label{fig:schem}

\vspace{0.5em}
\footnotesize \textit{Note}: The survey was fielded from January 2025 to June 2025. The response rates for the computational and general samples were about 24\% and 11\%, respectively. The weighting symbols represent, from left to right, gender identity, location, and subsample.
\end{figure}

\subsection{Measures}

Our survey focuses on who uses GenAI, its use cases, reasons for and concerns over using GenAI, and sources of optimism and trust in the technology. See the replication repository for information about accessing the survey questionnaire and data, which includes variables not explored in the main text.

\subsection{Analytic Plan}

One particular interest of our work is to describe differences between scholars who self-report using computational methods and those who do not. To do this, we use an array of univariate, bivariate, and multivariate descriptive data visualizations. We also use regression models to analyze the extent trust and optimism are associated with levels of GenAI use and familiarity. All descriptive analyses and regressions are rake-weighted with weight-adjusted standard errors and confidence intervals \citep{lumley2017fitting}. See Appendix~\ref{app_barcharts} for our findings as percentages over individual respondent categories (e.g., $X\%$ of respondents stated they ``somewhat agree'' with $Y$). Recall that our sample includes sociologists and their collaborators. To simplify, we refer to this population collectively as \textit{scholars}. 

\newpage
\section{Findings}

\subsection{Who Uses GenAI?}

\begin{figure}[hbt!]
\centering\includegraphics[width=1\linewidth]{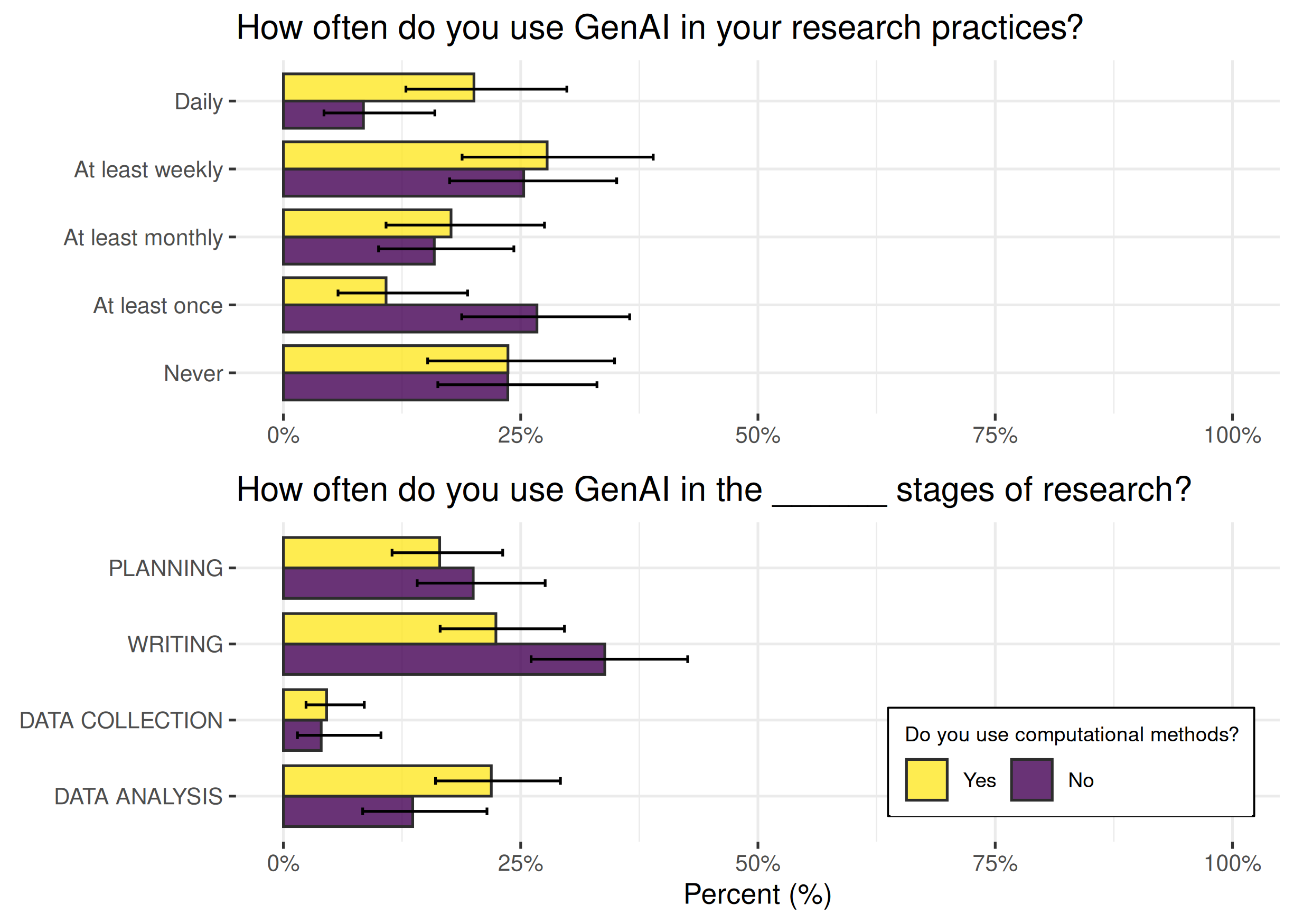}
\caption{Distribution of GenAI Usage Frequency}
\label{fig:fig1_use}

\vspace{0.5em}
\footnotesize \textit{Note}: Error bars are 95\% confidence intervals. Top panel $n=$ 394 after listwise deletion and bottom panel $n=$ 334 after removing respondents who responded having never used GenAI and listwise deletion.
\end{figure}

How many scholars use GenAI and for what purpose(s)? The top panel of Figure \ref{fig:fig1_use} shows the distribution of the frequency with which GenAI is used in research practices. Computational scholars use GenAI with higher frequency than non-computational scholars, but the differences are generally small. For computational scholars, ``Weekly'' is the most frequent response (27.8\%; 18.8\% $\leq \mu \leq$ 39\%), but with ``Never'' a close second (23.7\%; 15.2\% $\leq \mu \leq$ 34.9\%), while non-computational say ``At least once'' the most (26.7\% compared to 10.8\%) with ``Weekly'' (25.3\%; 17.5\% $\leq \mu \leq$ 36.5\%) a close second.

\FloatBarrier

\subsection{GenAI Use Cases}

Of those who have used GenAI in their research at least once, how do they use it? The second panel of Figure \ref{fig:fig1_use} shows the distribution of GenAI use at various stages of research---planning, writing, data collection, and analysis---separated by non-computational and computational scholars. The percentages sum to 100\% per research stage, but with the never-users included in the calculation and removed from the visualization. As the bar chart shows, computational scholars are slightly more likely to use GenAI in analysis tasks (21.9\%; 16\% $\leq \mu \leq$ 29.2\%) than non-computational scholars (13.6\%; 8.4\% $\leq \mu \leq$ 21.4\%). Outside this research stage, however, the differences are marginal. Importantly, while the never-users are not shown, they make up the largest group for each research stage---in other words, most scholars do not use GenAI in any of the research stages as defined by the survey. Nonetheless, among those who have used GenAI, writing is the most common research stage where GenAI comes into play for both non-computational and computational scholars. In the Appendix (Figure \ref{fig:fig1_use_app} in Appendix~\ref{app_barcharts}) we break down each research stage into specific tasks, such as translating one's own writing or explaining statistical outputs.\footnote{Generally, we find very few differences, the most significant of which is in the analysis stage. Specifically and unsurprisingly, computational scholars are using GenAI to help with coding -- writing (20.8\%) and debugging (21.8\%). Non-computational scholars are also somewhat more likely to use GenAI for grammar and spelling assistance (32.6\%) and translation (24.8\%). That said, both computational and non-computational scholars most often use GenAI to help with writing tasks. Though analysis is tied with writing for computational scholars.}

Of course, the options provided do not encompass all possible tasks. Some noted in an open-ended response field that they use GenAI for generating paper titles, creating standardized images for survey stimuli, or automating emails related to conducting research. Furthermore, some respondents indicated that groupings of tasks could be disaggregated, e.g., annotating text may be distinguished from classification.

\FloatBarrier

\subsection{Reasons for Using GenAI}

\begin{figure}[hbt!]
\centering\includegraphics[width=.95\linewidth]{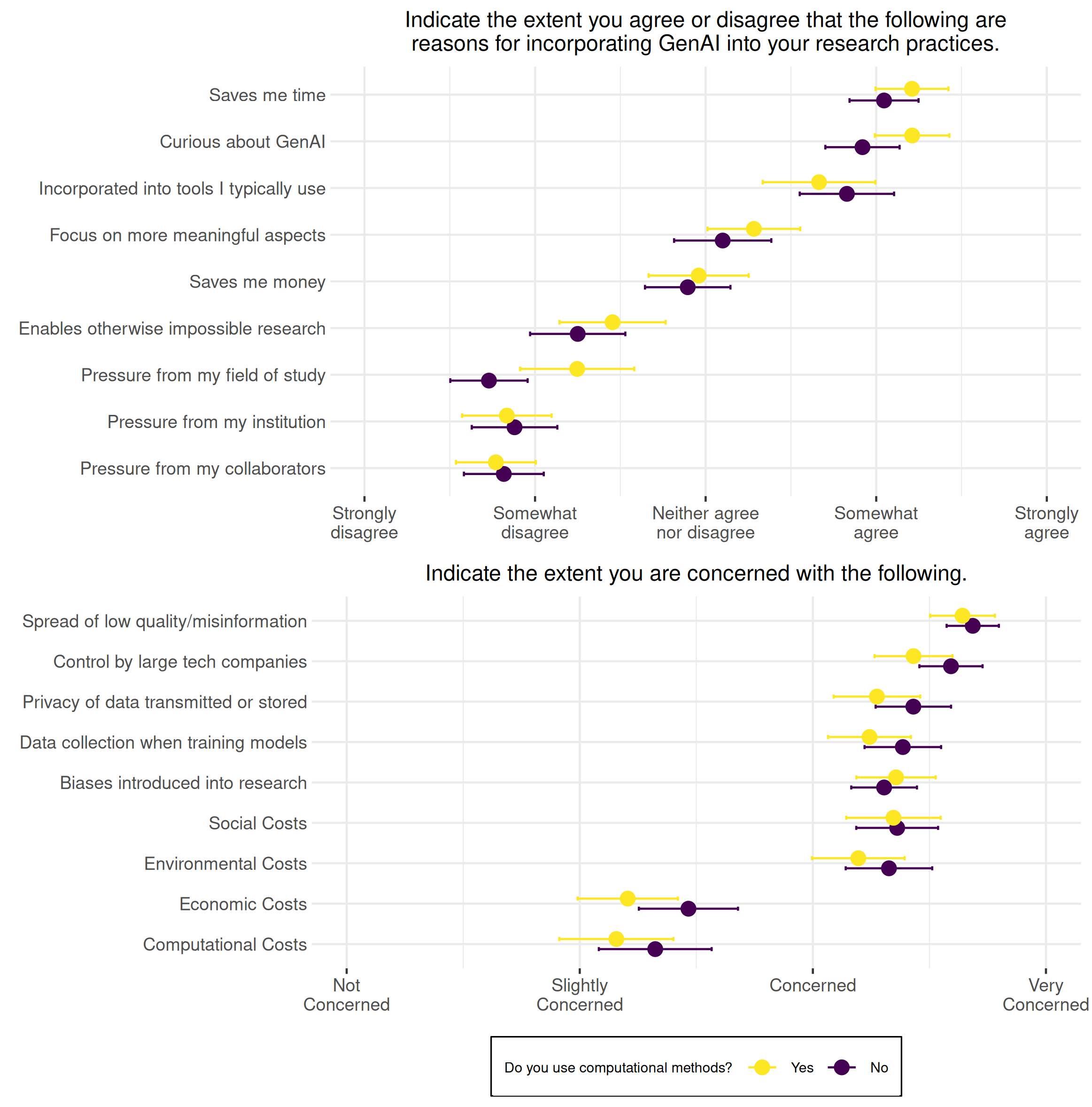}
\caption{Reasons for, and Concerns about, Using GenAI in Research}
\label{fig:fig2}

\vspace{0.5em}
\footnotesize \textit{Note}: Points are means. Error bars are 95\% confidence intervals. Top panel: $n=318$ after removing respondents who responded having never used GenAI in their research, as well as listwise deletion. Bottom panel: $n=376$ after listwise deletion.
\end{figure}

Of those who use GenAI in their research, what are the reasons for doing so? To answer this question, we rely on a combination of closed-ended items and representative selections from open-ended responses. The top panel of Figure \ref{fig:fig2} visualizes the mean responses for possible reasons, split between computational and non-computational scholars. There was general agreement among scholars that use GenAI because they say it saves them time and satisfies their curiosity. Scholars also report using GenAI because tools they typically use are now incorporating GenAI. As one respondent remarked:

\begin{quote}
I only use GenAI because Google has normalized AI summaries in their standard searches. I otherwise do not use GenAI in my work.
\end{quote}

Scholars are more ambivalent on whether it allows them to focus on more meaningful aspects of research or whether it saves them money, and do not think it enables otherwise impossible research. 

External pressures may also be driving reasons to adopt GenAI in research. As one respondent noted:

\begin{quote}
I feel like people are going to be using it more and more, and I do not want to be left behind. I am worried about that my research process may become too `slow' or obsolete.
\end{quote}

Although this respondent was not alone, the top panel of Figure \ref{fig:fig2} suggests that pressures from collaborators, the field, or institutions are generally low for computational and non-computational scholars alike. Additionally, several open-ended responses from scholars who were not native English speakers noted that GenAI helps them be more confident in their writing (though not for direct translation of their work). As before, there are not many meaningful distinctions between non-computational and computational scholars.

\FloatBarrier

\subsection{Concerns about GenAI}

Among all respondents, what possible costs or risks of GenAI are concerning? When asked about general ``costs'' of GenAI, scholars express the most concern about its environmental and social impacts (bottom panel of Figure \ref{fig:fig2}). Although non-computational scholars were, on average, more worried about nearly all costs, the differences with computational scholars are not significant. When asked about a series of more specific consequences, respondents expressed the most concern about information quality and the level of control held by large tech companies over GenAI tools. Still, scholars were also highly concerned about the privacy of their data, data collection practices used in training GenAI, and GenAI biasing research practices. On all questions, there are, again, no strong differences between non-computational and computational scholars.

Overall, our items indicate that there is a high level of concern about various aspects of GenAI; of the nine questions we asked, seven were rated as being between ``Concerned'' and ``Very Concerned'' on average. To capture possible nuance regarding respondents' perceptions of risks, we also asked about their GenAI concerns in an open-ended question. For instance, one respondent wrote that they were ``wholly against the use of GenAI tools. They have no place in social science practice.'' Below, we complement the aggregate findings with further examples that we found to be representative of general themes in the open-ended responses.

One pattern we found is that scholars do not just worry about the spread of low quality content/misinformation, but that GenAI may lead to a general reduction in critical thinking. One respondent summarizes this common sentiment: 

\begin{quote}
I am deeply concerned that the use of GenAI and other AI tools will result in significant compromises in research, including false information, researchers following false leads, reduction in human critical thinking, and reduction in the value of human knowledge/critical thinking/thought. At this stage, GenAI and AI tools pose a threat to the integrity of knowledge that is factual, thoughtful, reflective, nuanced, and critical.
\end{quote}

Respondents also noted that GenAI may undermine the development of tacit knowledge that is important for sound research---for example, one wrote that ``[t]here are key cognitive scaffolds that come from practice which might be undermined by easy use [of GenAI] for simple tasks.'' They also voiced concerns that GenAI may exacerbate the current trend to publish more \citep{Warren2019-rd}. As a respondent explains, 

\begin{quote}
Ideally, GenAI would be used to relieve some of the workload that comes with the `publish or perish' culture. However, if the faster output it helps to generate just increases expectations for that much more output, then it may increase the number of studies done but won't be of much personal benefit to stressed and overworked researchers/students.
\end{quote}

Finally, respondents noted that current training, policies, and ethical standards regarding GenAI are lagging in the field. For instance, one respondent stated that ``[p]eople are not trained, or are poorly trained, in GenAI and do not understand how to use it appropriately. This leads to widespread misuse of this tool.'' In fact, some suggest that GenAI itself may be a hindrance to such training: 

\begin{quote}
GenAI makes computational analytical tools accessible to more researchers. However, I’m concerned that it might undermine the training in computational methods, as students may no longer feel motivated to develop solid programming skills or a deep understanding of computational theories.
\end{quote}

This lack of a deeper understanding as to how GenAI works may create fertile ground for unscrupulous behavior, as noted by one respondent: 
\begin{quote}
I think there are a lot of overly credulous academics, the ones getting the most money for generative AI in research right now, who genuinely think LLMs are magic... they take a totally unscientific approach to what these objects are ontologically and how they work mechanically.
\end{quote}

\FloatBarrier

\subsection{Optimism and Trust}

Finally, we measure the optimism and trust about the future of GenAI for research. Scholars are generally more optimistic that GenAI will continue to improve over the next 2 to 3 years (40.2\% of all scholars ``Somewhat agree'' or ``Strongly agree'')---though there is ambivalence about whether this will have a net positive effect on the field or whether the current advantages of GenAI outweigh the drawbacks for the field. Furthermore, scholars generally believed that GenAI outputs cannot be trusted (only 4.5\% of all scholars ``Somewhat agree'' or ``Strongly agree'' that GenAI can be trusted, see also Figure \ref{fig:bar7_8} in Appendix~\ref{app_barcharts}).

Skepticism was also a common theme in the open-ended responses: 

\begin{quote}
I treat GenAI as an inexperienced, mediocre RA. I don't trust it, but use it to do simple things I can easily check
\end{quote}

and 

\begin{quote}
I think we're at the early stages here. I think it can sometimes be useful to bounce ideas off, but often it returns quite derivative feedback, and I have not found these tools very useful for searching for articles.
\end{quote}

We used rake-weighted generalized linear regression models to further probe heterogeneity in trust and optimism.\footnote{We also ran a series of rake-weighted ordered logistic regression models as a robustness check given the ordinal nature of the outcome variables. They are consistent with the simpler linear model results reported here. These alternative models are in Table \ref{tab:reg2} in Appendix~\ref{reg2}. Adjusted predictions for these models are visualized in Figure \ref{fig:app}.} Specifically, we examine whether scholars who are more familiar with GenAI are more likely to trust GenAI, believe that GenAI will improve in the next 2 to 3 years, or believe that GenAI will have a net positive impact on their field in the next 2 to 3 years, and if those effects might be moderated by their use level.

Our outcome variables were the Likert-type ``future improvement,'' ``net positive,'' and ``trust'' questions (see Figure \ref{fig:bar7_8}), respectively. All three variables range from ``Strongly Disagree'' = 1 to ``Strongly Agree'' = 5. The two primary covariates are (1) a binary variable equal to 1 if the respondent reports using GenAI in their research at least weekly and a 0 otherwise, and (2) a ``familiarity'' composite score computed as the sum across three variables: a question asking respondents to indicate the extent they are familiar with GenAI, understand how GenAI works, and are confident in their ability to use GenAI.\footnote{Each of these three questions was measured using the same Likert scale as the one used for the outcome variables. The three items have high reliability (Cronbach's $\alpha=0.87$) and are unidimensional according to a principal component analysis ($\lambda_1$ = 2.37).} We control for gender identity using a binary variable (1 = not a cis-man and 0 = cis-man).

\begin{figure}[hbt!]
\centering\includegraphics[width=1\linewidth]{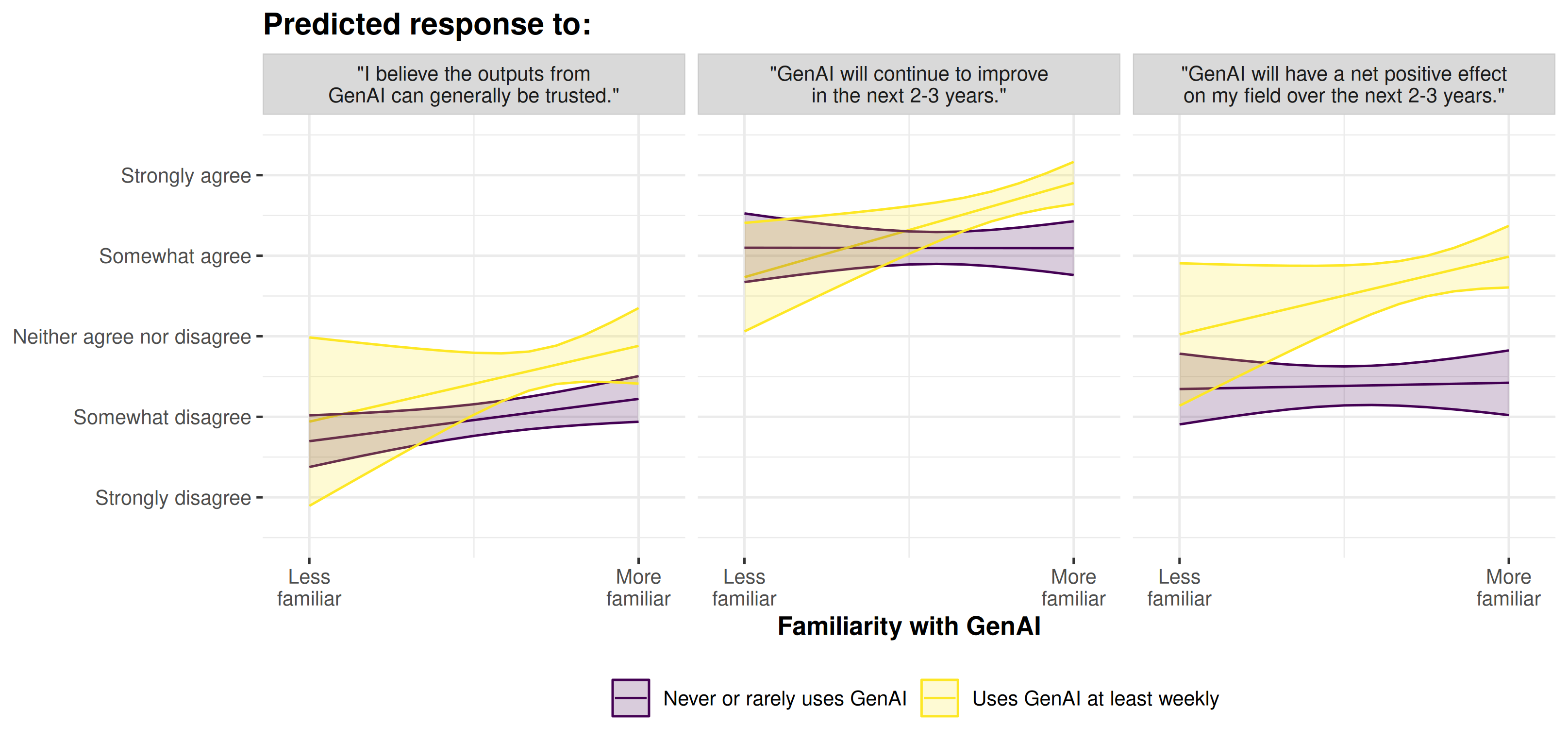}
\caption{Adjusted Predictions of GenAI Trust, Future Improvement, and Net Positive for Research}
\label{fig:fig9}

\vspace{0.5em}
\footnotesize \textit{Note}: Error ribbons represent 95\% confidence intervals. Predictions based on $n=$ 411, 407, and 409, respectively, after listwise deletion. Gender identity is held at the modal value (``cis-man'').
\end{figure}

The regression results are in Table \ref{tab:reg1} in Appendix~\ref{reg1} with unstandardized estimates. A model with only familiarity as a predictor of trust (not shown in Table \ref{tab:reg1}) had a very small positive effect ($\hat{\beta}$ = 0.078 with $t$ = 3.660). However, use level is not a significant moderator of this effect (see Model 1 in Table \ref{tab:reg1}). Use level is also not a significant moderator for the net positive outcome variable but is for the future improvement variable. The regression results are summarized with adjusted predictions in Figure \ref{fig:fig9}, where gender identity is held at its modal value (cis-man). 

As predictions show in the furthest left panel, trust is generally low regardless of use-level or familiarity. Optimism that GenAI will improve, on the other hand, is generally high regardless of use level or familiarity---with regular users showing a bit more variation by familiarity level (as suggested by the significant interaction term). However, beliefs that this will result in positive effects for sociology are somewhat higher for regular users (about neutral to somewhat agree) than for non-regular users (somewhat disagree). Use level again does not moderate the relationship between familiarity and perceptions of net positive effects, but both active and non-active users tend to be somewhat ambivalent on this issue. Part of this ambivalence can be decoupled from how useful or accurate GenAI may actually be, as it may lead to the delegation of key research tasks. Respondents commented that this arrangement might, in the long term, ``produce increasingly poorer researchers'', or undermine critical thinking leading to a ``very slippery slope.'' 

We note that across the three models (see Table \ref{tab:reg1}), there is a consistent negative association with gender identity. Controlling for familiarity and use levels, non cis-men are less likely than cis-men to trust GenAI output, think that it will improve in the short term, or agree that GenAI will have a net positive effect on their field in the near future. Though not statistically significant in these models, the negative estimates are in line with surveys of the general U.S. adult population which found that women, nonbinary, and transgender groups hold more negative attitudes about AI \citep{haimson2025ai}.

\FloatBarrier

\section{Discussion}

We presented findings from a survey asking sociologists and their collaborators how they use (or do not use) GenAI in their research and their attitudes toward the technology more broadly. Thereby, we provided an assessment of generative AI's role in contemporary sociological research. This assessment is not conclusive, but instead a snapshot that captures a specific moment in a rapidly evolving context. We hope that our survey will serve as a baseline for continued efforts to monitor how and whether GenAI diffuses across the discipline. In closing, we point out important fault lines and questions for the future.

First, our survey suggests that scholars are currently not feeling pressures to adopt GenAI in their own research. The confluence of capital, labor, technology, and politics that formed the GenAI ecosystem has already caused disruption and pressures in a variety of other fields. As the landscape of availability changes---i.e., GenAI being incorporated into more applications, such as email, word processors, qualitative data analysis suites or integrated development environments---and as universities and other organizations implement GenAI policies, it will be important to monitor the extent to which sociologists perceive being expected to use these new tools.

Second, we do not yet know to what extent sociologists might see GenAI as a threat to their sociological expertise. Sociological work may be challenged by GenAI as even the specialties most removed from computational methodology, namely sociological theory, are implicated.\footnote{Incidentally, the most elite sociology departments in the US specialize in theory \citep{elder2025status}.} More generally, GenAI threatens knowledge work by broadening the availability of required expertise, weakening occupational closure mechanisms, and reducing bargaining power. Such tensions are further exacerbated by the increased dependence on science and technology in tandem with a decreasing trust in scientists \citep{eyal2019crisis}. This routinization of knowledge production may eventually allow for the displacement of well-paid professionals for ``workers paid a fraction of their wage, flood[ing] the market with inferior products, and impos[ing] unreasonable and punishing working conditions on those'' using GenAI \citep[39]{Bender2025-wa}, c.f., \citep{Nelson2025-zb}. Social scientists might come to feel these tensions more acutely than they do now, given that GenAI can present \textit{convincing or even compelling} arguments about human society and social behavior, even if these are inaccurate or entirely fabricated.

Third, attitudes and concerns about GenAI may change considerably if scholars become more familiar with and knowledgeable about the technology. Our study does not allow us to draw any causal inferences, but some of our results could at least be interpreted as hinting in this direction: familiarity was (weakly) associated with trust in GenAI outputs, and those who use GenAI more frequently are generally more optimistic about its effects on the discipline. Sociologists using GenAI usually have interactional expertise as opposed to contributory expertise \citep{collins2019rethinking}, that is, they know how to use these tools and how to trigger desired outputs through carefully crafted prompts. If more sociologists find ways to make GenAI useful for their work, this will likely have an effect on attitudes and raise new or exacerbate existing concerns. However, use in this sense is not necessary for gaining a deeper understanding of GenAI. That is, there may be sociologists with rich contributory expertise, but who remain skeptical of this technology and refuse to use it in their research.

Finally, our study suggests that there is wide heterogeneity in both usage and attitudes. Some scholars use GenAI daily for a variety of tasks and are optimistic about its effects on the discipline. Many others have never used it and see no role or at most a very limited one for it in sociological research. Surprisingly, these differences did not fall along the lines of division that we had anticipated (computational/non-computational scholars). Nevertheless, our results point to a significant challenge for the discipline: establishing norms on GenAI's proper use (if any) in the research process. We hope that our study makes a small contribution to addressing this challenge by mapping out how GenAI is currently being used. Future studies might focus more directly on the normative considerations surrounding GenAI use than we did. This could help us locate a consensus on what we should consider appropriate and inappropriate uses of GenAI in sociological research.

\FloatBarrier
\bibliographystyle{plain}
\bibliography{genai}  

\FloatBarrier
\pagebreak

\appendix

\section{Appendix: Survey Invitation Primer on GenAI}\label{app0}

\appendix

\section{Example Recruitment Materials}\label{app0}

\subsection{Invitation Email}

Dear [REDACTED], 

I hope this email finds you well. Together with my colleagues [REDACTED], I am conducting a survey on how sociologists and their collaborators use and think about generative AI.

We are reaching out to you specifically because you recently published an article in a sociology journal. We would like to invite you to participate in our short online survey. We are interested in your thoughts, regardless of whether you have experience using generative AI. By participating, you will shape discussions on generative AI in sociology and help us understand its role, if any, in the sociological toolkit. Most people take between 5 and 10 minutes to complete the survey. All responses will be kept confidential.
 
To take this survey, please follow this personalized link: [REDACTED]

Or copy and paste the URL below into your internet browser: [REDACTED]

The above link is unique to your email address. If you prefer to take part anonymously, please use this link: [REDACTED]

\subsection{Generative AI Explainer}

For the sake of this survey, we will use GenAI to refer to generative models that produce content (e.g., GPT-4), as well as specific applications or platforms built on top of them (e.g., ChatGPT).

There is wide variation in how GenAI might be used in social scientific research. Its potential applications range from writing/outlining, transcription, translation, internet search, summarization and literature review, to data annotation/labeling, coding/debugging, and simulating human behavior and interactions.

Some specific GenAI tools include:
\begin{compactitem}
    \item OpenAI's ChatGPT browser-based chat interface 
    \item OpenAI's Whisper transcription service
    \item Google's Gemini, providing, for example, summaries in search results
    \item Perplexity's conversational academic search engine
    \item GitHub's Copilot, providing automated code writing
    \item Microsoft's Copilot, generating PowerPoint slides
    \item Meta's Llama models, which can run local chat interfaces
    \item Mistral's models, which can run local chat interfaces
    \item Stability's Stable Diffusion, which generates images from text
\end{compactitem}

Some models (GPT-4) run on remote servers and require an internet connection to access (e.g., through ChatGPT or OpenAI's API), however other models (e.g., Llama) can be operated "locally" on a desktop or laptop without the need for an internet connection.

Considering all these applications and models, as well as others we may not have listed, how often do you use GenAI in your research practices?

\pagebreak

\section{Sampling Methods}\label{app3}

\subsection{Design}
A scholar's expertise, contributory or interactional \`a la Collins and Evans \citep{collins2019rethinking}, in relation to GenAI is likely to shape both usage and attitudes. Specifically, we expected that \textit{computational} scholars would bring a potentially unique perspective on GenAI as it has emerged out of the same computational techniques that many such scholars use (i.e., machine learning and NLP). However, comparatively, this group is a very small subset of the broader sociological community. We therefore make an effort to both identify these scholars from our larger sampling frame and oversample them. In addition to having a better understanding of how these technologies operate, we also expected that computational scholars will have a higher response rate due to intrinsic interest and will be more likely to use GenAI in some capacity. Furthermore, based on prior surveys of (generative and non-generative) AI usage in academia, we expected that men will also be more likely to respond (possibly, in part, because they are overrepresented among computational scholars). We therefore design the survey to allow us to adjust for some parameters of our population. Although we attempt to control for such bias, inferences to the entire field of sociology must be made with caution. See the schematic representation of our sampling strategy in the main text(Figure \ref{fig:schem}).

\begin{table}[hbt!]
\caption{50 Journals Used to Select Sociology Authors}
\begin{tabular}[t]{>{\itshape}l >{\itshape}l}
\toprule
Acta Sociologica & Social Currents\\
American Journal of Cultural Sociology & Social Forces\\
American Journal of Sociology & Social Networks\\
American Sociological Review & Social Problems\\
Annual Review of Sociology & Social Psychology Quarterly\\
\addlinespace
British Journal of Sociology & Social Science Research\\
City \& Community & Society and Mental Health\\
Criminology & Socio-Economic Review\\
Cultural Sociology & Sociological Forum\\
Current Sociology & Sociological Inquiry\\
\addlinespace
Demography & Sociological Methodology\\
Environmental Sociology & Sociological Methods and Research\\
Ethnic and Racial Studies & Sociological Perspectives\\
European Sociological Review & Sociological Science\\
Gender \& Society & Sociological Theory\\
\addlinespace
International Journal of Sociology & Sociology\\
Journal for the Scientific Study of Religion & Sociology Compass\\
Journal of Health and Social Behavior & Sociology of Education\\
Journal of Marriage and Family & Sociology of Health \& Illness\\
Journal of Mathematical Sociology & Sociology of Race and Ethnicity\\
\addlinespace
Poetics & Sociology of Religion\\
Rationality and Society & Socius\\
Rural Sociology & Symbolic Interaction\\
Sexualities & Work and Occupations\\
Social Compass & Work Employment and Society\\
\bottomrule
\end{tabular}
\label{tab:journals} 
\end{table}

We build our sampling frame using a bibliometric approach with the Clarivate Web of Science (WoS) database. We collected all articles published in 50 sociology journals (see Table \ref{tab:journals}) in the last 5 years (2020-2025).\footnote{As of 10 February 2025.} This yielded a total of roughly 18,600 authors. Note that we include both sociologists (in title) and their collaborators.

Next, we classified an author in our sampling frame as ``computational" if they authored an article that references specific terms (see Table \ref{tab:terms}) in their title, abstract, or keywords. Our intent was to be broad in our boundary drawing, and thus capture the diversity of perspectives within the field of computational sociology. For additional precision, we also manually read a selection of titles and abstracts to ensure that we were capturing papers that engage with computational ideas and methods. In total, this yielded 985 authors who have written 405 computational articles in the last five years in sociology journals. To contact the authors, we combined the email addresses from the WoS records with manually searching for addresses posted publicly on faculty pages. Where an email ``bounced" (i.e., rejected by the receiving server), one attempt was made to find an accurate address and resend the email. This resulted in a total of 879 authors in our computational sample with valid email addresses.

From the remaining authors (who had not authored a computational paper as defined above), we randomly selected 2,200 individuals. We again used the email addresses from the WoS records and manually searched for those with missing addresses, and attempted to find more up-to-date email addresses in the case of bounces. This resulted in 1,943 invitations sent to individuals in our non-computational sample with valid email addresses. 

We began fielding the survey at the end of January 2025, and continued until the end of June 2025. Of the roughly 3,000 individuals invited to complete the survey, 219 completed the survey from the computational sample (\textasciitilde24.1\% response rate) and 214 completed the survey from the general sample (\textasciitilde11\% response rate), for a total of 433 respondents.

\begin{table}[hbt!]
\caption{Terms Used to Select ``Computational" Articles}
\centering
\begin{tabular}{p{5.5cm} p{8.5cm}}
\toprule
Category                    &  Term                     \\ \hline
Generative AI               & generative AI, genAI, chatGPT, OpenAI, GPT-2, GPT-3, 
                              GPT-4, DALL-E, Llama, MIXTRAL, MISTRAL, Stable Diffusion, 
                              prompt engineering, large language models, LLMs     \\ \hline
Machine Learning            & machine learning, artificial intelligence,  
                              deep learning, reinforcement learning, neural network\\ \hline
Natural Language Processing & natural language processing, NLP, text analysis, 
                              text mining, text as data                             \\ \hline
General                     & computational, data science, big data, 
                              algorithm, agent based modeling     \\                                                                                       
\bottomrule
\end{tabular}
\label{tab:terms} 
\end{table}

\FloatBarrier

\subsection{Weighting}

By inferring the gender of authors using their names and/or pronouns (in authors' biographies), we estimate that roughly 39\% of the total sampling frame are men. However, this jumps to 64\% for our computational subsample. Furthermore, using the base domains\footnote{In the few cases where the email address used a generic domain like gmail.com, we confirmed the location of the primary affiliation using their faculty biographies.} of email addresses, we inferred authors' locations. While roughly 24\% of authors in the total sampling frame were in the United States, this jumps to about 42\% in the computational subsample. Thus, oversampling on computational authors, as well as likely self-selection, shifted our sample toward men in the United States (see Table \ref{tab:demos}). We therefore use weights to correct for these discrepancies. 

Specifically, we use raking \citep{debell2009computing}. This procedure forces univariate distributions of the \textit{self-reported} variables gender and location to equal the \textit{inferred} population parameters in Table \ref{tab:demos} (in the ``Total" column). We also condition on the size of our computational subsample (i.e., those publishing a computational article in a sociology journal in the last 5 years), given that we oversampled computational scholars and also found that this subsample had a higher response rate (24\%) compared to the general subsample (10\%). All together, then, our raking weights force the marginal distributions of gender, US-based, and subsample type to equal the first column of Table \ref{tab:demos}. For the sake of raking, we use a binarized version of the respective variables (cis-man or not, US-based or not, and computational sample or not).

\begin{table}[]
\caption{Population, Sample, and Respondent Demographics}
\centering
\begin{tabular}{p{3.5cm} c c c c}
\toprule
                     &      Total    &    General  &  Computational  &    Respondents   \\
                     &    (Inferred) &  (Inferred) &  (Inferred)     &  (Self-Reported) \\ \hline
 Gender (\% Men)     &   39.40\%     &   38.1\%    &  64.3\%         &  54.3 \%         \\
 Location (\% US)    &   24.40\%     &   23.4\%    &  42.5\%         &  46.7 \%         \\
 Computational (Self-Reported) &    4.95\%   &   - &  -              &  50.6\%          \\
\bottomrule
\end{tabular}
\label{tab:demos} 
\end{table}

\FloatBarrier

The sum of the rake weights equals the total number of respondents in the total sample ($n=433$). Each respondent-specific weight quantifies ``how much" of the sample they represent as a function of their gender identity, location, and subsample. The maximum weight is 3.28 for non cis-men not based in the U.S. and who \textit{have not} published a computational article in the last 5 years; the minimum weight is 0.04 for cis-men based in the U.S. who \textit{have} published a computational article in the last 5 years.

In Table \ref{tab:rake}, we present the raw proportion for a range of descriptive variables, and their rake-weighted counterparts (note that each variable is presented as a binary response). Importantly, we are inferring to the population defined by the 50 sociology journals, which may not represent the field of sociology by all definitions. After weighting across these three population parameters, we can infer that roughly 30\% of the population contributing to the sociological literature considers themselves a quantitative (as opposed to qualitative or mixed-methods) scholar; 39\% identify as cis-men; 36\% are native English speakers; 34\% are tenured professors; and 60\% consider themselves racially white.\footnote{Although our sample is quite international, the gender and race proportions are nevertheless closely aligned with the demographic breakdowns of the 2024 membership of the American Sociological Association: 39.2\% cis-men and 54.4\% white (https://www.asanet.org/diversity-equity-inclusion/dei-at-asa/asa-membership/current-membership-2024/).}

\begin{table}[]
\caption{Raw Proportions and Rake-Weighted Proportions on Select Variables}
\centering
\begin{tabular}[t]{lcc}
\toprule
                          & Proportion (Raw) & Proportion (Rake) \\
\midrule
Gender (\% Cis-Men)       & 52.9\% & 39.40\%\\
Location (\% US)          & 45.7\% & 24.40\%\\
Computational             & 54.7\% & 34.90\%\\
Sociologist               & 71.8\% & 63.00\%\\
Quantitative              & 40.0\% & 30.40\%\\
Language (\% English)     & 47.1\% & 36.70\%\\
Race/Ethnicity (\% White) & 62.8\% & 60.90\%\\
Position (\% Tenured Prof.) & 34.4\% & 34.07\%\\
\bottomrule
\end{tabular}
\label{tab:rake} 
\end{table}

\FloatBarrier

\section{Descriptive Bar Charts}\label{app_barcharts}

The following Figure \ref{fig:fig1_use_app} breaks down the research stages by individual tasks. For each panel the percentages sum to 100\% per specific use case with the never-users omitted from the visualization.

\begin{figure}[hbt!]
\centering\includegraphics[width=1\linewidth]{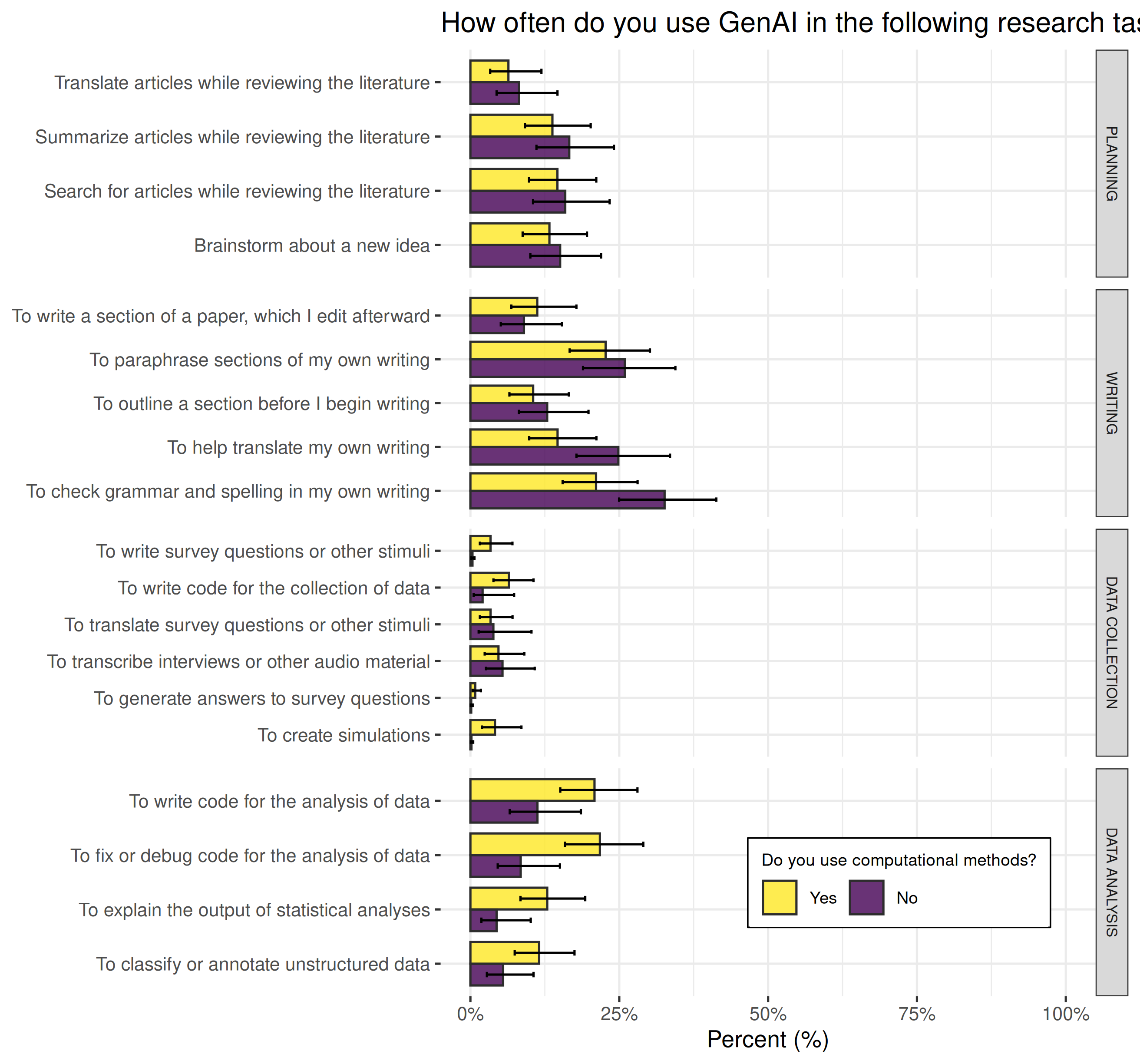}
\caption{Scholars GenAI Use Frequency by Specific Research Tasks}
\begin{tablenotes}
\item \footnotesize \textit{Note}: $n=$ 334, 216, 244, 143, and 200 for overall, planning, writing, data collection, and analysis, respectively, after removing respondents who responded having never used GenAI, and listwise deletion. Error bars are 95\% confidence intervals.
\end{tablenotes}
\label{fig:fig1_use_app}
\end{figure}

\FloatBarrier

\begin{figure}[hbt!]
\centering\includegraphics[width=1\linewidth]{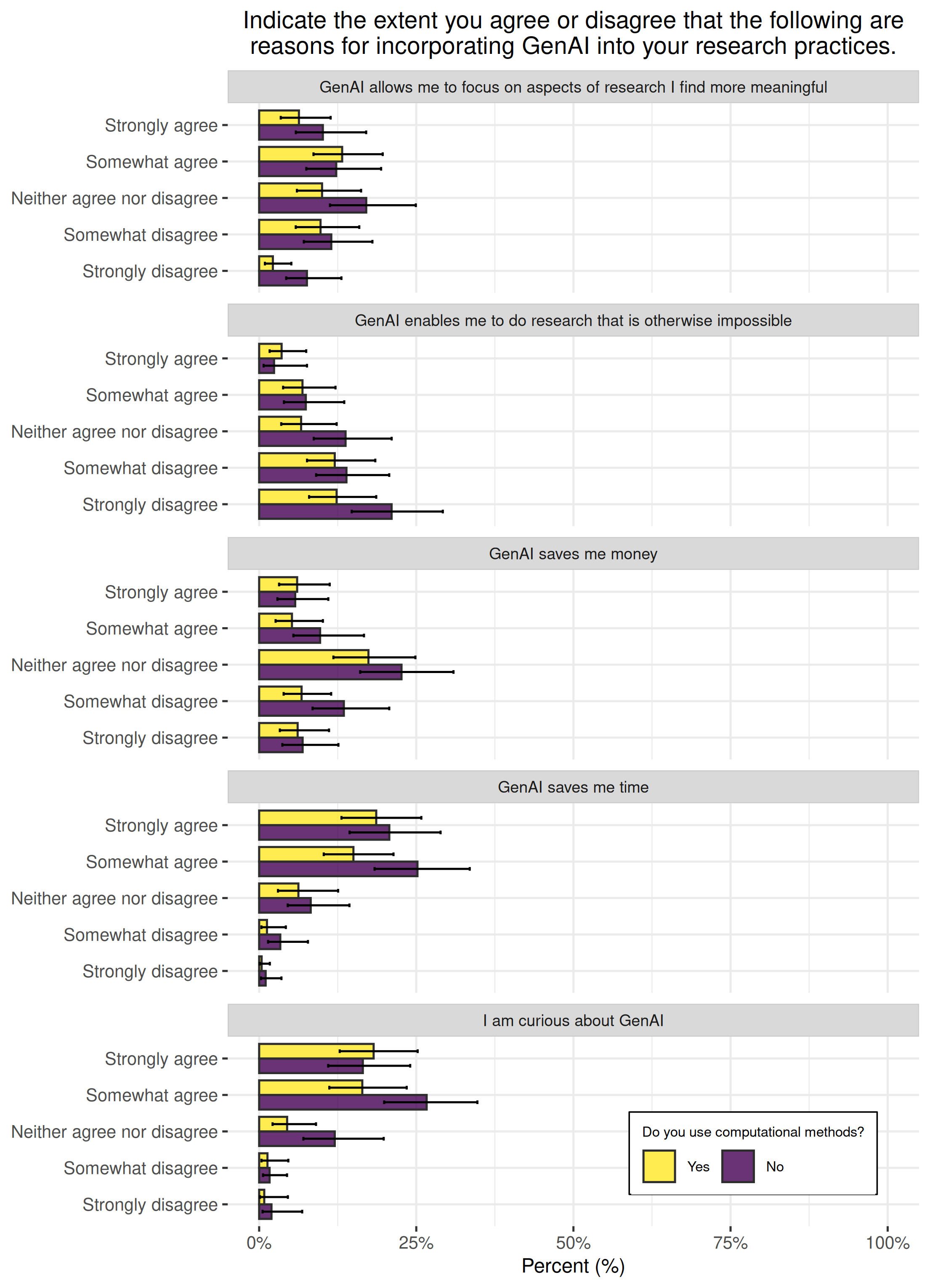}
\caption{Reasons for Using GenAI in Research}
\begin{tablenotes}
\item \footnotesize \textit{Note}: $n=310$ after listwise deletion. Error bars are 95\% confidence intervals.
\end{tablenotes}
\label{fig:bar3}
\end{figure}

The next figures visualize the distributions for all questions across each Likert-type category. The percentages sum to 100\% per reason---meaning that, for example in \ref{fig:bar3}, about 17\% of scholars who answer the question about GenAI affording them the opportunity to do more meaningful research are non-computational scholars who neither disagree nor agree with this reason.

\begin{figure}[hbt!]
\centering\includegraphics[width=1\linewidth]{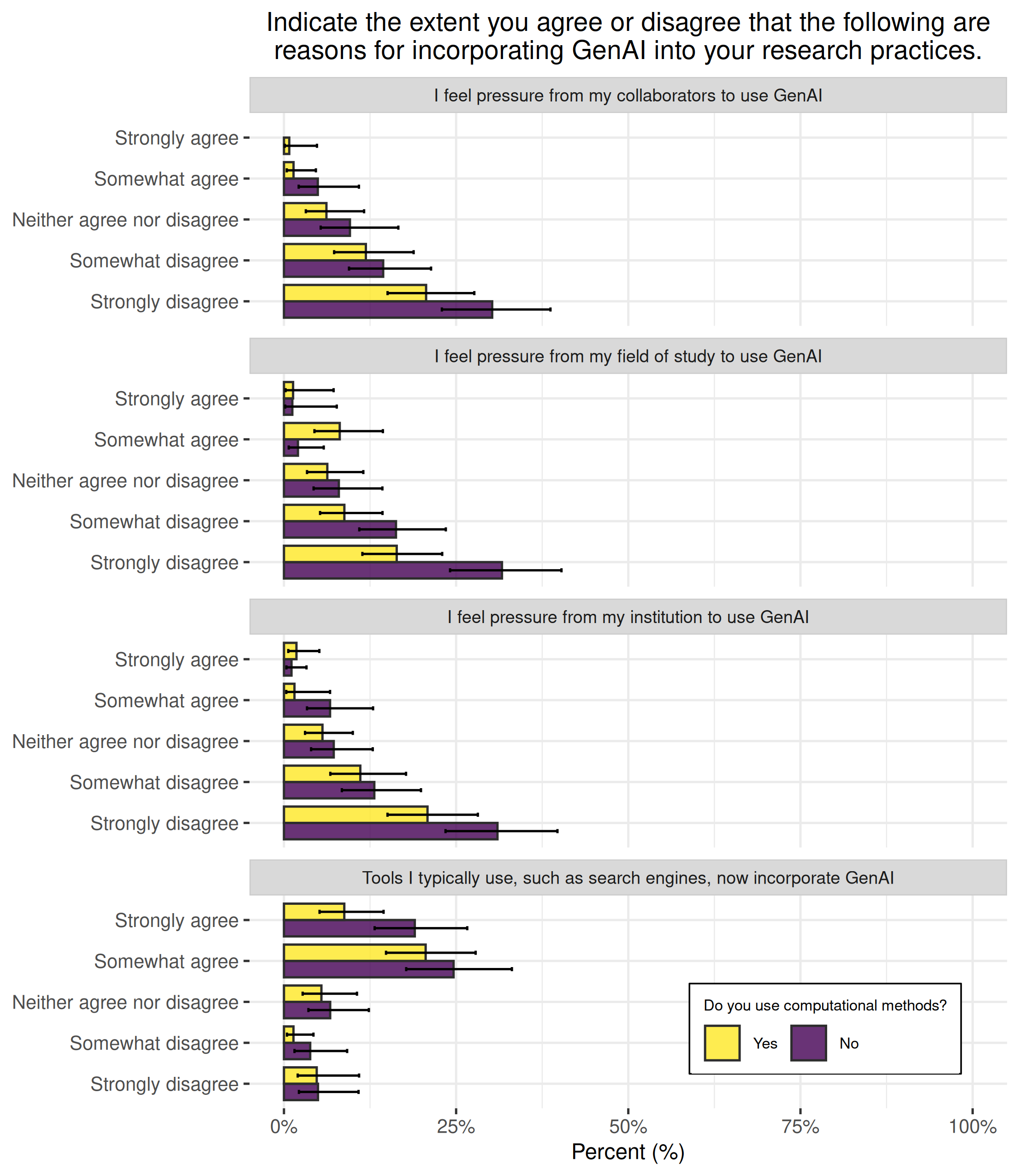}
\caption{Reasons for Using GenAI in Research: External Pressures}
\begin{tablenotes}
\item \footnotesize \textit{Note}: $n=308$ after removing respondents who responded having never used GenAI in their research, as well as listwise deletion.
\end{tablenotes}
\label{fig:bar4}
\end{figure}

\begin{figure}[hbt!]
\centering\includegraphics[width=1\linewidth]{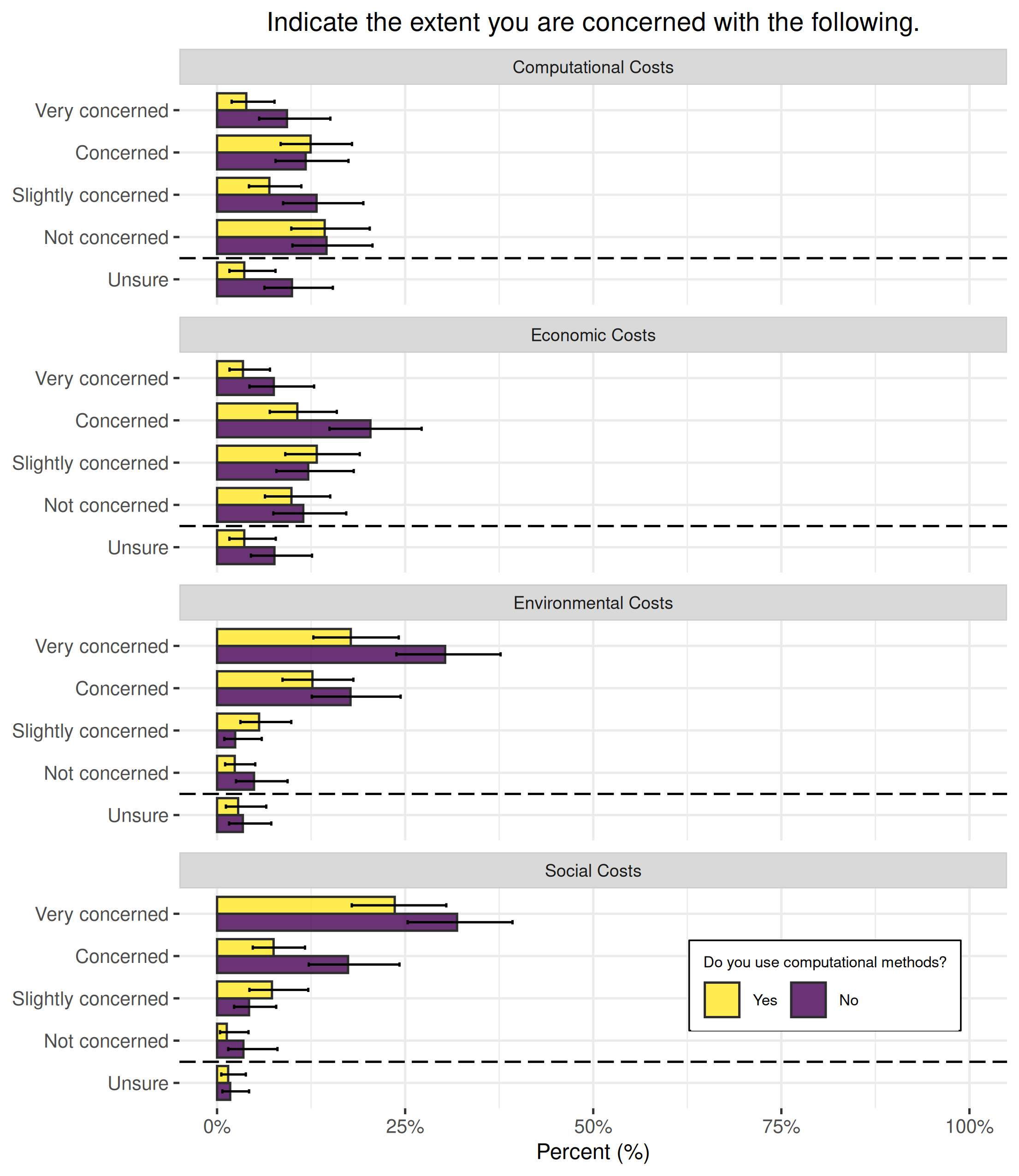}
\caption{Costs of Using GenAI in Research}
\begin{tablenotes}
\item \footnotesize \textit{Note}: $n=392$ after listwise deletion. Error bars are 95\% confidence intervals.
\end{tablenotes}
\label{fig:bar5}
\end{figure}

\begin{figure}[hbt!]
\centering\includegraphics[width=1\linewidth]{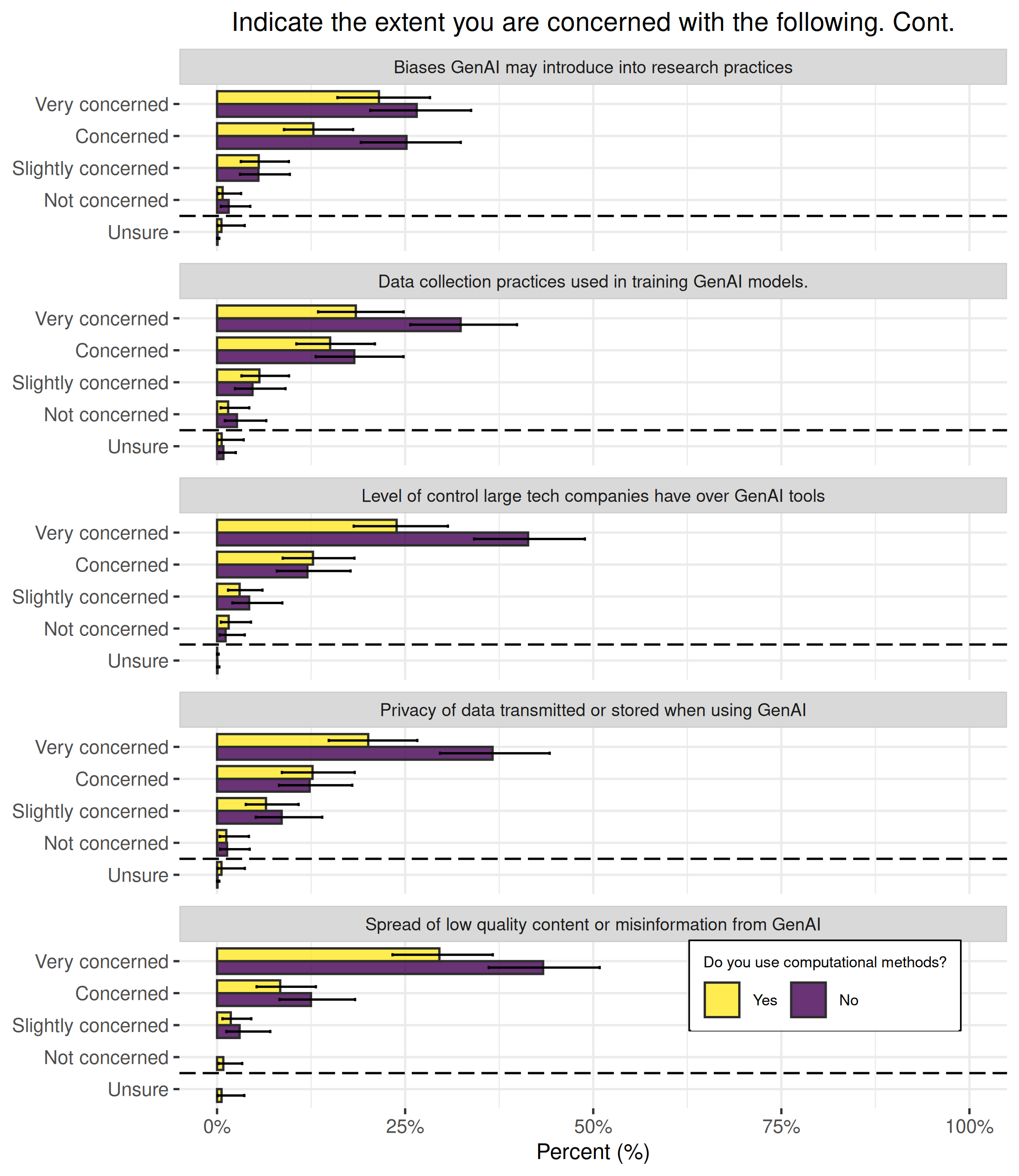}
\caption{Concerns about Using GenAI in Research}
\begin{tablenotes}
\item \footnotesize \textit{Note}: $n=393$ after listwise deletion. Error bars are 95\% confidence intervals.
\end{tablenotes}
\label{fig:bar6}
\end{figure}

\begin{figure}[hbt!]
\centering\includegraphics[width=1\linewidth]{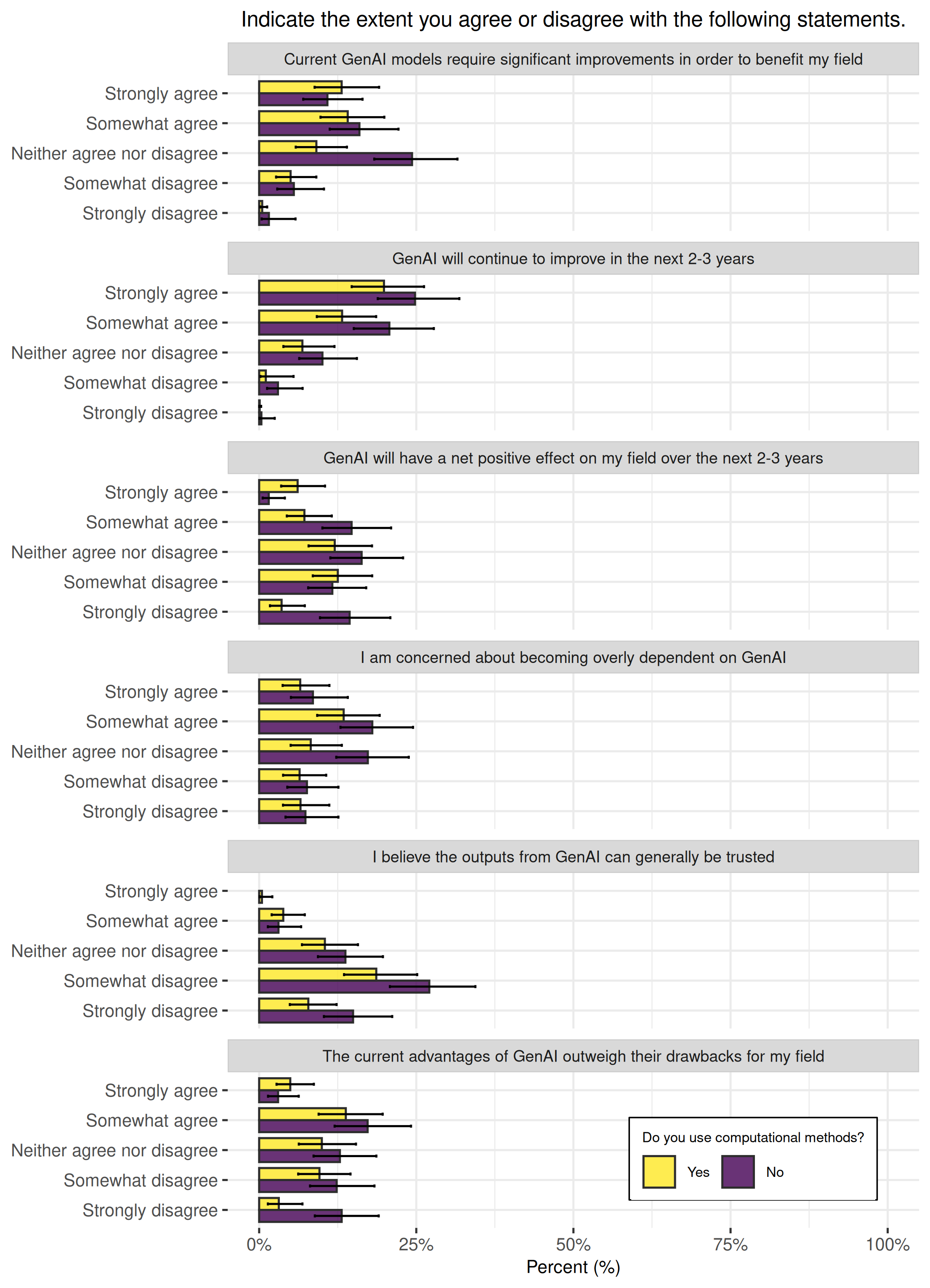}
\caption{Distrust and Optimism about Using GenAI in Research}
\begin{tablenotes}
\item \footnotesize \textit{Note}: $n=392$ after listwise deletion. Error bars are 95\% confidence intervals.
\end{tablenotes}
\label{fig:bar7_8}
\end{figure}

\FloatBarrier

\section{Regression Model}\label{reg1}

\begin{table}[hbt!]
\centering
\begin{threeparttable}
\caption{Weighted Generalized Linear Estimates of GenAI Trust, Future Improvement, and Net Positive for Research}

\begin{tabular}{lcccccc}
\toprule
 & \multicolumn{2}{c}{Trust} 
 & \multicolumn{2}{c}{Future Improvement} 
 & \multicolumn{2}{c}{Net Positive} \\
\cmidrule(lr){2-3}\cmidrule(lr){4-5}\cmidrule(lr){6-7}
 & $\hat{\beta}$ & $se_{\beta}$ 
 & $\hat{\beta}$ & $se_{\beta}$ 
 & $\hat{\beta}$ & $se_{\beta}$ \\
\midrule
Familiarity              & 0.044$^{*}$   & 0.020 & -0.000 & 0.027 & 0.007 & 0.029 \\
Use                      & 0.138         & 0.727 & -0.659 & 0.534 & 0.455 & 0.652 \\
Familiarity $\times$ Use & 0.035         & 0.064 & 0.098$^{*}$ & 0.045 & 0.074 & 0.056 \\
Not Cis-Man              & -0.015        & 0.125 & -0.144 & 0.115 & -0.113 & 0.150 \\
\\
Intercept                & 1.566$^{***}$ & 0.213 & 4.100$^{***}$ & 0.292 & 2.325$^{***}$ & 0.300 \\
\\
$N$                     & \multicolumn{2}{c}{411} 
                        & \multicolumn{2}{c}{407} 
                        & \multicolumn{2}{c}{409} \\
Pseudo-$R^2$            & \multicolumn{2}{c}{0.144} 
                        & \multicolumn{2}{c}{0.115} 
                        & \multicolumn{2}{c}{0.300} \\
AIC                     & \multicolumn{2}{c}{1025.377} 
                        & \multicolumn{2}{c}{982.277} 
                        & \multicolumn{2}{c}{1176.869} \\
$Dev_{model}$           & \multicolumn{2}{c}{286.636} 
                        & \multicolumn{2}{c}{264.318} 
                        & \multicolumn{2}{c}{426.790} \\
$Dev_{null}$            & \multicolumn{2}{c}{334.950} 
                        & \multicolumn{2}{c}{298.621} 
                        & \multicolumn{2}{c}{610.081} \\
\bottomrule
\end{tabular}

\begin{tablenotes}
\footnotesize
\item \textit{Note}: Coefficient estimates are unstandardized. Standard errors are design-adjusted \citep{lumley2017fitting}. 
$n=$ 411, 407, and 409, respectively, after listwise deletion. 
Reference category for ``Use'' is ``Not at least weekly.'' 
Reference category for ``Not Cis-Man'' is ``Cis-Man.'' 
$^{*}p<.05$, $^{**}p<.01$, $^{***}p<.001$ (two-tailed).
\end{tablenotes}

\label{tab:reg1}
\end{threeparttable}
\end{table}

\section{Alternative Model Specifications}\label{reg2}

\begin{table}[hbt!]
\centering
\begin{threeparttable}
\caption{Survey-Weighted Ordered Logistic Estimates of GenAI Trust, Future Improvement, and Net Positive for Research}

\begin{tabular}{lcccccc}
\toprule
 & \multicolumn{2}{c}{Trust} 
 & \multicolumn{2}{c}{Future Improvement} 
 & \multicolumn{2}{c}{Net Positive} \\
\cmidrule(lr){2-3}\cmidrule(lr){4-5}\cmidrule(lr){6-7}
 & $\hat{\beta}$ & $se_{\beta}$ 
 & $\hat{\beta}$ & $se_{\beta}$ 
 & $\hat{\beta}$ & $se_{\beta}$ \\
\midrule
Familiarity               & 0.096$^{*}$   & 0.048 & -0.006 & 0.062 & -0.000 & 0.052 \\
Use                       & 0.787         & 1.830 & -2.260 & 1.420 & 0.461 & 1.190 \\
Familiarity $\times$ Use  & 0.033         & 0.157 & 0.295$^{*}$ & 0.123 & 0.152 & 0.106 \\
Not Cis-Man               & -0.066        & 0.274 & -0.345 & 0.285 & -0.241 & 0.271 \\
\\
\textbf{Intercepts}       &               &       &        &       &        &       \\
$Y \leq$ Strongly Disagree               & -0.039$ $ & 0.529 & -5.490$^{***}$ & 1.140 & -1.180$^{*}$ & 0.564 \\
$Y \leq$ Somewhat Disagree               & 2.110$^{***}$ & 0.544 & -3.120$^{***}$ & 0.719 & 0.191 & 0.541 \\
$Y \leq$ Neither Agree nor Disagree      & 3.960$^{***}$ & 0.576 & -1.310 & 0.685 & 1.730$^{***}$ & 0.513 \\
$Y \leq$ Somewhat Agree                  & 6.120$^{***}$ & 0.849 & 0.405 & 0.677 & 3.390$^{***}$ & 0.580 \\
\\
$N$                      & \multicolumn{2}{c}{411} 
                         & \multicolumn{2}{c}{407} 
                         & \multicolumn{2}{c}{409} \\
$Dev_{model}$            & \multicolumn{2}{c}{1005.255} 
                         & \multicolumn{2}{c}{888.167} 
                         & \multicolumn{2}{c}{1141.478} \\
$Dev_{null}$             & \multicolumn{2}{c}{1069.968} 
                         & \multicolumn{2}{c}{947.538} 
                         & \multicolumn{2}{c}{1286.297} \\
$\chi^2_{null-model}$    & \multicolumn{2}{c}{64.712$^{***}$} 
                         & \multicolumn{2}{c}{59.371$^{***}$} 
                         & \multicolumn{2}{c}{144.818$^{***}$} \\
\bottomrule
\end{tabular}

\begin{tablenotes}
\footnotesize
\item \textit{Note}: Standard errors are design-adjusted \citep{lumley2017fitting}. 
Estimates are expressed in log odds. $n=$ 411, 407, and 409, respectively, after listwise deletion. 
Reference category for ``Use'' is ``Not at least weekly.'' 
Reference category for ``Not Cis-Man'' is ``Cis-Man.'' 
$^{*}p<.05$, $^{**}p<.01$, $^{***}p<.001$ (two-tailed).
\end{tablenotes}

\label{tab:reg2}
\end{threeparttable}
\end{table}

\begin{figure}[hbt!]
\centering
\includegraphics[width=\textwidth]{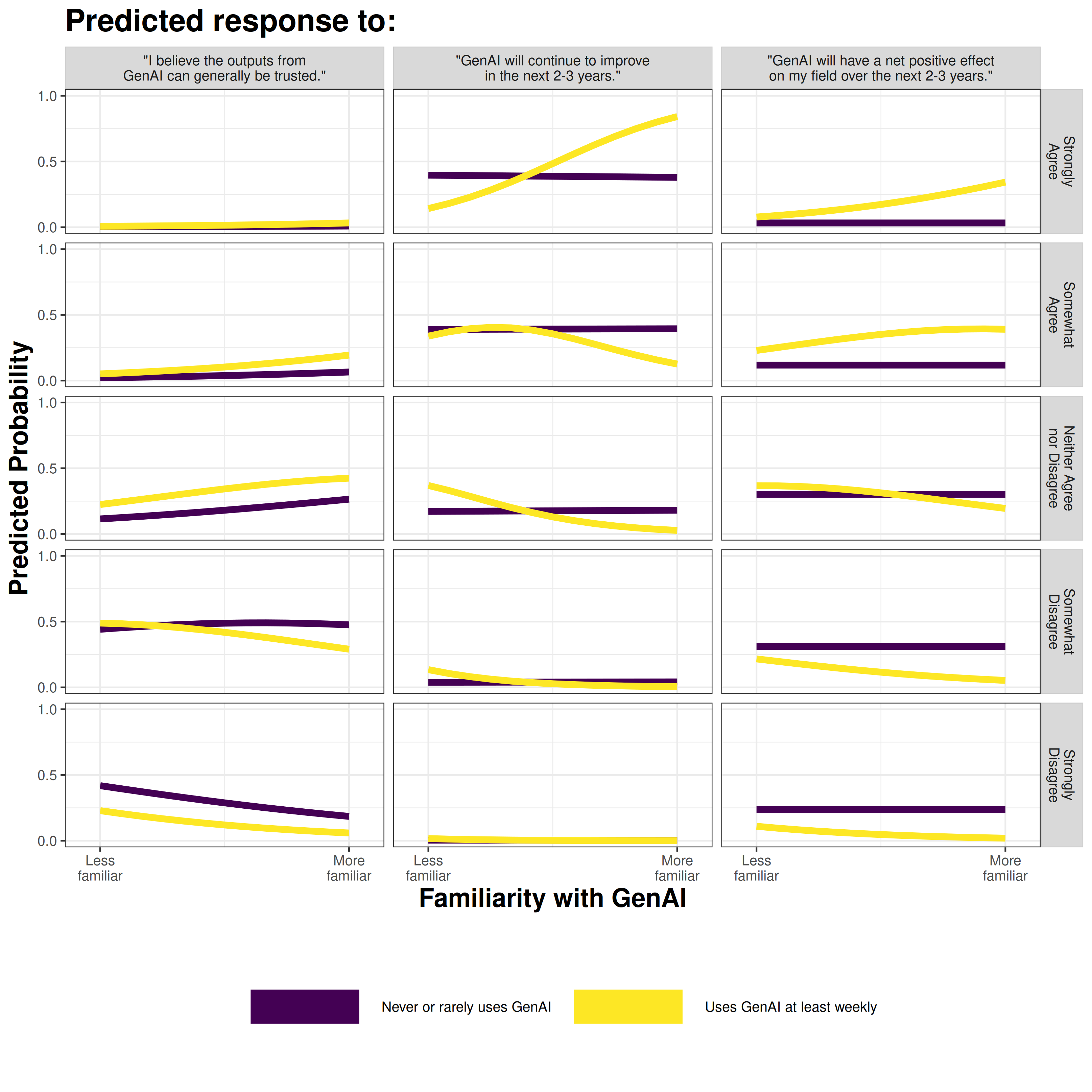}
\caption{Adjusted Predictions of GenAI Trust, Future Improvement, and Net Positive for Research from Survey-Weighted Ordered Logistic Regressions}
\begin{tablenotes}
\footnotesize
\item \textit{Note}: Coefficient estimates are unstandardized. Predictions based on $n=$ 411, 407, and 409, respectively, after listwise deletion. Gender identity control is held at the modal value (``cis-men'').
\end{tablenotes}
\label{fig:app}
\end{figure}

\end{document}